# Lessons from early Earth: UV surface radiation should not limit the habitability of active M star systems

Jack T. O'Malley-James⋆ and L. Kaltenegger
*Carl Sagan Institute, Cornell University, Ithaca, NY 14853, USA*

**ABSTRACT**

The closest potentially habitable worlds outside our Solar system orbit a different kind of star than our Sun: smaller red dwarf stars. Such stars can flare frequently, bombarding their planets with biologically damaging high-energy UV radiation, placing planetary atmospheres at risk of erosion and bringing the habitability of these worlds into question. However, the surface UV flux on these worlds is unknown. Here we show the first models of the surface UV environments of the four closest potentially habitable exoplanets: Proxima-b, TRAPPIST-1e, Ross-128b, and LHS-1140b assuming different atmospheric compositions, spanning Earth-analogue to eroded and anoxic atmospheres and compare them to levels for Earth throughout its geological evolution. Even for planet models with eroded and anoxic atmospheres, surface UV radiation remains below early Earth levels, even during flares. Given that the early Earth was inhabited, we show that UV radiation should not be a limiting factor for the habitability of planets orbiting M stars. Our closest neighbouring worlds remain intriguing targets for the search for life beyond our Solar system.

## 1 INTRODUCTION

M stars are the most common type of star in the Galaxy and make up 75 per cent of the stars in the solar neighbourhood. They are also excellent candidates to search for terrestrial planets in the liquid water orbital distance range (or the so-called habitable zone HZ), due to the high frequency of rocky planets in the HZs of these stars (Gaidos 2013; Dressing & Charbonneau 2013, 2015). Our nearest known small and likely rocky HZ planets – Proxima-b, Ross 128b, TRAPPIST-1e, -f, -g, LHS 1140-b (Anglada-Escudé et al. 2016; Gillon et al. 2016; Dittmann et al. 2017; Bonfils et al. 2018) – all orbit M stars. Proxima Centauri, a cool, active M6V dwarf only 1.3 parsec from the Sun, harbours a planet in its HZ with a minimum mass of 1.3 Earth masses that receives about 65 per cent of Earth's solar flux (Anglada-Escudé et al. 2016). At 3.4 parsec from the Sun, the planet Ross 128b, with a minimum mass of about 1.4 Earth masses, orbits in the HZ of its cool, inactive M4V dwarf star (Bonfils et al. 2018). The TRAPPIST-1 planetary system of seven transiting Earth-sized planets around a cool, moderately active M8V dwarf star, which has several (three to four) Earth-sized planets in its HZ, is only about 12 parsec from the Sun (e.g. Gillon et al. 2017; O'Malley-James & Kaltenegger 2017; Ramirez & Kaltenegger 2017). The planet LHS 1140b orbits in the HZ of its cool, likely inactive M4.5V dwarf star, with a measured rocky composition based on its radius of 1.4 Earth radii and mass of 6.7 Earth masses (Dittmann et al. 2017). These four planetary systems already provide an intriguing set of close-by potentially habitable worlds for the search for life beyond our own Solar system.

Planets in M star systems may face potential barriers to habitability as a result of their host star's activity compared to Earth, especially if their host star is active and produces strong UV radiation (see e.g. discussion in Scalo et al. 2007; Tarter et al. 2007; Shields, Ballard & Johnson 2016; France et al. 2016; Kaltenegger 2017; Loyd et al. 2018). Furthermore, the high EUV/X-ray and charged particle fluxes associated with active, flaring M stars could place the atmospheres and water inventories of their HZ planets at risk over time (Vidotto et al. 2013; Garraffo, Drake & Cohen 2016; Kreidberg & Loeb 2016; Ribas et al. 2016; Turbet et al. 2016; Airapetian et al. 2017; Dong et al. 2017; Garcia-Sage et al. 2017; Kopparapu et al. 2017; Lingam & Loeb 2017; Barnes et al. 2018; Goldblatt 2018; Meadows et al. 2018), eroding them, especially as the close proximity of planets to their host star in the HZs of cool stars can cause planetary magnetic fields to be compressed by stellar magnetic pressure, reducing a planet's ability to resist atmospheric erosion by the stellar wind (Lammer et al. 2007; See et al. 2014). Therefore planets that receive high doses of UV radiation are generally considered to be less promising candidates in the search for life (see e.g. Buccino, Lemarchand & Mauas 2006) given that when UV radiation is absorbed by biological molecules, especially nucleic acids, harmful effects such as mutation or inactivation can result, with shorter UV wavelengths having the most damaging effects (see e.g. Kerwin & Remmele 2007). Certain radiation-tolerant species have demonstrated an ability to survive full solar UV in space exposure experiments (e.g. Sancho et al. 2007; Onofri et al. 2012); however, they achieve this by entering a dormant state. Therefore, although life may be able to survive on highly UV-irradiated surfaces like this, it would likely not be able to actively metabolize and complete a life cycle.

⋆ E-mail: jomalleyjames@astro.cornell.edu

**Table 1.** The properties of the closest four star systems that are currently known to harbour potentially habitable planets. These systems represent our current best close-by targets in searches for life beyond the Solar system. Unless otherwise stated, stellar data were obtained from Anglada-Escudé et al. (2016) [Proxima Centauri], Bonfils et al. (2018) [Ross 128], Gillon et al. (2016) [TRAPPIST-1], and Dittmann et al. (2017) [LHS 1140].

|  | Proxima Centauri | Ross 128 | TRAPPIST-1 | LHS-1140 |
|---|---|---|---|---|
| **Stellar type** | M5.5/6 V | M4 V | M8 V | M4.5 V |
| **Distance (pc)** | 1.3 | 3.38 ± 0.006 | 12.1 ± 0.4 | 12.5 ± 0.4 |
| **Luminosity ($L_\odot$)** | 0.00155 ± 0.00006 | 0.00362 ± 0.00039 | 0.000524 ± 0.000034 | 0.00298 ± 0.00021 |
| **$T_{eff}$ (K)** | 3050 ± 100 | 3192 ± 60 | 2559 ± 50 | 3131 ± 100 |
| **Age (Gyr)** | 4.8[a] | ≥ 5 | 3-8[b] | > 5 |
| **Habitable zone (AU)** | 0.03 – 0.09 | 0.05 – 0.13 | 0.02 – 0.05 | 0.04 – 0.11 |
| **Habitable zone Angular separation (milliarcsec)** | 23 – 69 | 14 – 38 | 1.6 – 4.1 | 3.2 – 8.8 |
| **Planet(s):** | *Proxima Centauri b* | *Ross 128b* | *TRAPPIST-1d, 1e, 1f, 1g* | *LHS-1140b* |
| **Orbital distance (AU)** | 0.0485 | 0.05 | 0.022, 0.029, 0.038, 0.046 | 0.0875 |
| **Angular separation (milliarcsec)** | 37 | 15 | 1.8, 2.4, 3.1, 3.8 | 7 |

[a]Bazot et al. (2016).
[b]Luger et al. (2017).

M stars remain active for longer periods of time compared to the Sun (see e.g. West et al. 2011). Flares from active M stars can increase the surface UV flux on their planet in the HZ by up to two orders of magnitude for up to several hours for the most active M stars modelled (Segura et al. 2010; Tilley et al. 2019). However, several teams have made the case that planets in the HZs of M stars can remain habitable, despite periodic high UV fluxes (see e.g. Heath et al. 1999; Buccino et al. 2007; Scalo et al. 2007; Tarter et al. 2007; Rugheimer et al. 2015; O'Malley-James & Kaltenegger 2017). Note that recent studies suggest that high UV surface flux may even be necessary for prebiotic chemistry to occur (see Ranjan & Sasselov 2016; Rimmer et al. 2018).

Here we explore the UV surface environments for the best close-by targets for the search for life among our neighbouring worlds. We show the biologically relevant UV-A (320–400 nm), UV-B (280–320 nm), and UV-C (100–280 nm) fluxes that reach the ground on these worlds, for a range of atmospheric conditions for Earth-like atmospheres from present-day analogues to eroded atmospheres, as well as considering anoxic atmospheres without ozone, which provides shielding from high-energy UV radiation for present-day Earth. We compare our results to UV surface level models for Earth throughout its geological evolution from 3.9 billion years ago to present-day Earth (see Rugheimer et al. 2015). Section 2 describes our models, Section 3 provides our results and discussion.

## 2 METHODS

Table 1 shows the parameters for the closest potentially habitable worlds beyond our Solar system and their four host stars. Comparing the angular separation [$\theta$(arcsec) = $a$(AU)/$d$(pc), where $a$ = planet semimajor axis and $d$ = distance to star system] for the HZ and the known planets within the HZ for each host star with the inner working angle (IWA) for a telescope, which describes the minimum angular separation at which a faint object can be detected around a bright star. This shows which planets can be remotely detected and resolved in the near future. For example the 38 m diameter Extremely Large Telescope (ELT) should have an IWA of 6 mas when observing in the visible region of the spectrum (assuming $\theta_{IWA} \approx 2(\lambda/D)$, where $\lambda$ is the observing wavelength and $D$ is the telescope diameter). Thus the planets Proxima-b, Ross-128b, and LHS-1140b can be resolved with the ELT for future in-depth characterization of these worlds.

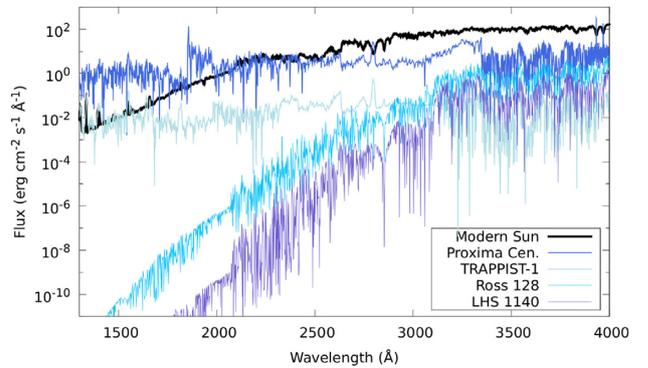

**Figure 1.** Stellar UV model spectra in the HZs of a flaring Proxima Centauri (blue), a flaring TRAPPIST-1(light blue), Ross-128 (turquoise), and LHS-1140 (lilac). The modern solar UV flux (black) is plotted for comparison. The UV fluxes in the HZs of active M stars can equal, or exceed, the solar UV flux that reaches present-day Earth as a result of the proximity of M star HZs to the host star.

We follow the methods of O'Malley-James & Kaltenegger (2017), using the stellar flux models shown in Fig. 1 and the planetary orbital parameters (see Table 1) of each planet modelled, assuming (i) an Earth-like atmosphere with 1 bar surface pressure; (ii) an eroded, lower density Earth-like atmosphere at 0.5 bar and 0.1 bar surface pressure, and (iii) an anoxic (trace levels of $O_2$; 10 per cent $CO_2$) 1 bar atmosphere analogous to the early Earth's atmosphere before the rise of oxygen (following Kaltenegger, Traub & Jucks 2007). Note that for the eroded atmosphere scenarios we maintain Earth-like mixing ratios of the component atmospheric gases, which results in lower column depths due to the lower total atmospheric mass.

We use a coupled climate-photochemistry code developed for rocky exoplanets, EXO-PRIME (see Kaltenegger & Sasselov 2009 for details), to model the UV surface fluxes on the planets we consider here under different atmospheric scenarios. It iterates between a 1D climate and a 1D photochemistry code to calculate the atmosphere transmission of UV fluxes to the ground of Earth-sized planets and calculates the corresponding atmosphere composition and observable spectrum. The climate code (originally developed by Kasting, Holland & Pinto 1985) utilizes a two-stream approximation (see Toon et al. 1989), which includes multiple scattering by atmospheric gases, in the visible/near-infrared to calculate the shortwave fluxes. Four-term, correlated-k coefficients parametrize the absorption by $O_3$, $H_2O$, $O_2$, and $CH_4$ (see Pavlov et al. 2000; Kopparapu et al. 2014). The photochemistry code (originally developed by Kasting et al. 1985 and updated by Segura et al. 2005 and references therein) solves for 55 chemical species linked by 220 reactions using a reverse-Euler method. For the anoxic atmosphere cases we use a 1D photochemical model for high-$CO_2$/high-$CH_4$ terrestrial atmospheres (see Pavlov et al. 2000; Kharecha et al. 2005; Segura et al. 2007; Rugheimer et al. 2015 and references therein). Note that these models do not account for scattering, which would reduce the amplitude of all the predicted surface fluxes; however the relative differences between the different scenarios, model planet and early Earth, would remain unchanged. Our stellar input spectra are based on publicly available UV spectra in active (flaring) and quiescent states where available (see e.g. France et al. 2013; Youngblood et al. 2016), covering a wavelength range of 1150–3350 Å. We combine them with PHOENIX models for a given star's characteristics. The compositions of our model atmospheres, temperature profiles and atmospheric $H_2O$, $O_3$, and $CH_4$ profiles with altitude, derived from these simulations are shown in Fig. 2. The eroded atmosphere planet models show higher $H_2O$ mixing ratios compared to the 1 bar atmospheres as a result of the decreased boiling point of water at lower pressures, however the overall column depth is lower, due to the lower overall atmospheric mass. UV fluxes shortwards of 240 nm, which are higher in our active star models, drive the photodissociation of oxygen, driving $O_3$ production. Ozone mixing ratios for high stellar UV activity are similar to present-day Earth's. For eroded atmospheres, the maximum $O_3$ concentration occurs at lower altitudes due to decreased atmospheric pressure. The $O_3$ mixing ratio is similar in the 1 bar and eroded atmospheres, but the overall $O_3$ column depth is lower in eroded atmospheres due to the lower total atmospheric mass. $CH_4$ mixing ratios are higher in these models than on present-day Earth because M stars emit lower fluxes in the 200–300 nm range that drives $CH_4$ photodissociation (Segura et al. 2005), giving $CH_4$ a longer atmospheric lifetime than on present-day Earth. For an anoxic atmosphere, no significant ozone layer develops for either stellar model.

## 3 RESULTS AND DISCUSSION

### 3.1 Surface UV flux compared to Earth

At the location of their planets in the HZ, the active M stars in our sample – Proxima Centauri and TRAPPIST-1 – have UV fluxes that equal, or exceed, present-day solar UV flux during flare events, while the inactive M stars have UV fluxes orders of magnitude weaker (Fig. 1). A planet's atmospheric composition influences the surface UV environment, with thinner low-density atmospheres enabling more UV radiation to penetrate a planet's surface due to lower column-integrated number densities of UV-absorbing gases compared to denser atmospheres of the same composition. The surface UV flux estimates for the planets (Fig. 3) show that more high-energy UV radiation reaches the ground as atmospheric thickness and ozone levels decrease.

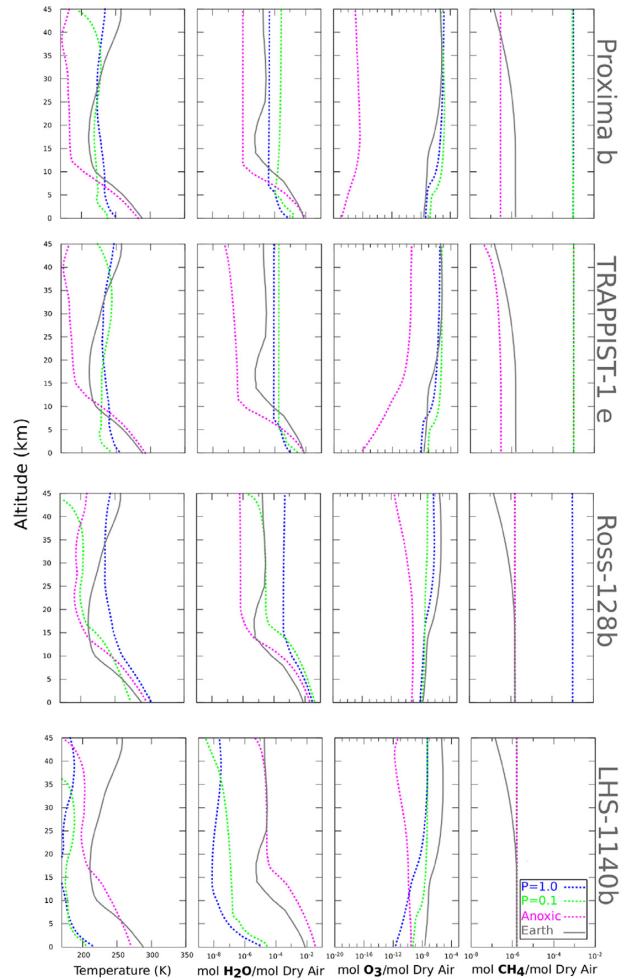

**Figure 2.** The chemical make-up for a range of model atmospheres for the planets investigated. The plots show profiles for temperature, $H_2O$, $O_3$, and $CH_4$ for a present-day Earth-like oxygen-rich atmosphere with surface pressures of 1.0 bar (blue), 0.1 bar (green), and an anoxic atmosphere (red) with a surface pressure of 1.0 bar. We plot present-day Earth–Sun profiles for comparison (grey). These differ from the 1.0 bar atmosphere models (blue) as a result of the different radiation environments our model planets experience from their M star hosts, compared to the solar radiation environment experienced by present-day Earth.

However, even though these planets in the HZs of active star systems receive higher UV fluxes than present-day Earth, their UV surface flux is lower than that of the early Earth 3.9 billion years ago (see Rugheimer et al. 2015) due to the lower top-of-atmosphere (TOA) UV flux for wavelengths larger than 200 nm from M stars compared to the Sun, even during flares.

In our atmosphere models, ozone filters out the most biologically harmful UV wavelengths shortwards of about 300 nm, as on present-day Earth, decreasing in effectiveness with decreasing ozone concentration. Shortwards of about 200 nm, absorption by atmospheric $CO_2$ filters out biologically harmful UV flux. Thus even if planets around active M stars have eroded atmospheres or do not contain ozone (anoxic), the resulting surface UV flux is still approximately an order magnitude lower than on the early Earth even for the planet orbiting the most active star in our sample, Proxima-b. The incident UV surface flux is about two orders of magnitude lower for TRAPPIST-1e. For planets orbiting inactive M stars the surface UV flux is even lower, which might result in a different concern for habitability, i.e. the question of whether such low UV surface levels could produce the macromolecular building blocks of life, assuming these require a certain minimum UV levels (Ranjan & Sasselov 2016; Rimmer et al. 2018).

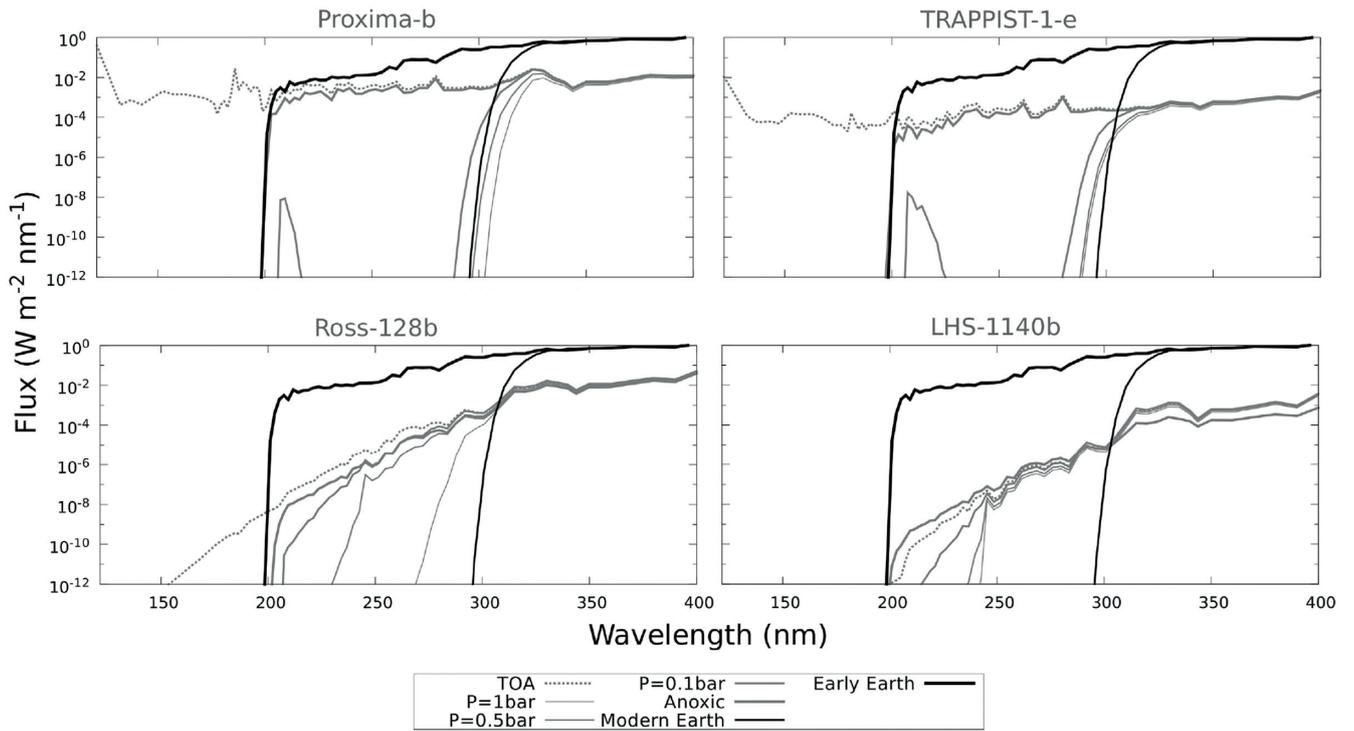

**Figure 3.** Modelled UV surface fluxes for present-day Proxima-b, TRAPPIST-1e, Ross-128b, and LHS-1140b for oxygen-containing atmospheres at pressures of 1.0 bar, 0.5 bar, and 0.1 bar, and a 1.0 bar anoxic atmosphere. Present-day and early Earth's UV surface flux is plotted for comparison.

### 3.2 Habitable surface UV environments and early Earth

The high-energy surface UV (UV-B and UV-C) cut-off occurs at shorter, more biologically harmful wavelengths for similar TOA UV fluxes as atmospheric pressure decreases and ozone concentration decreases. When UV radiation is absorbed by biological molecules, especially nucleic acids, harmful effects such as mutation or inactivation can result, with shorter UV wavelengths having the most damaging effects (see e.g. Kerwin & Remmele 2007).

Even though we cannot anticipate what kind of life could evolve on other worlds – assuming life could emerge on these worlds – we can explore the surface habitability of the closest potentially habitable planets in regard to known life on Earth. We use biological action spectra to show the relative biological effectiveness, i.e. a measure of biological damage at a given UV wavelength based on life as we know it, caused by a particular UV surface radiation environment (see two biological action spectra in panel A in Fig. 4). One shows the relative mortality rates at different UV wavelengths of the radiation-tolerant extremophile *Deinococcus radiodurans* (Setlow & Boling 1965; Calkins & Barcelo 1982). *Deinococcus radiodurans* is one of the most radiation-resistant organisms known on Earth (Rothschild & Mancinelli 2001). Therefore, we use this as a benchmark against which to compare the habitability of the different radiation models. This action spectrum compares the effectiveness of different wavelengths of UV radiation at inducing a 90 per cent mortality rate. It highlights which wavelengths have the most damaging irradiation for biological molecules: for example, the action spectrum in Fig. 4 shows that a dosage of UV radiation at 360 nm would need to be three orders of magnitude higher than a dosage of radiation at 260 nm to produce similar mortality rates in a population of this organism. Similarly, the second action spectrum describes the relative destruction of DNA molecules per unit time at different UV wavelengths (panel A, Fig. 4) (Voet et al. 1963; Diffey 1991). The convolution of the modelled surface UV fluxes in each atmosphere scenario with the action spectra show the relative survivability for *D. radiodurans* and DNA molecules under the different radiation regimes (Fig. 4).

Note that certain radiation-tolerant species have demonstrated an ability to survive full solar UV in space exposure experiments (e.g. Sancho et al. 2007; Onofri et al. 2012); however, they achieve this by entering a dormant state. Therefore, although life may be able to survive on highly UV-irradiated surfaces like this, it would likely not be able to actively metabolize and complete a life cycle. We use Proxima-b – the most UV-irradiated planet in our study – as an example (see Fig. 4).

Observations suggest that Proxima-b receives 30 times more EUV than present-day Earth, and 250 times more X-ray radiation (Ribas et al. 2016). Additionally, recent MOST observations found that low-energy white light flares occur as often as 63 times per day. Comparisons with similarly active stars suggest that large flares (with energies of $10^{33}$ erg) occur up to eight times per year (Davenport et al. 2016); a prediction supported by the recent detection of a superflare event (Pavlenko et al. 2017). Hence, the UV surface fluxes we model for Proxima-b during a flare could be occurring regularly.

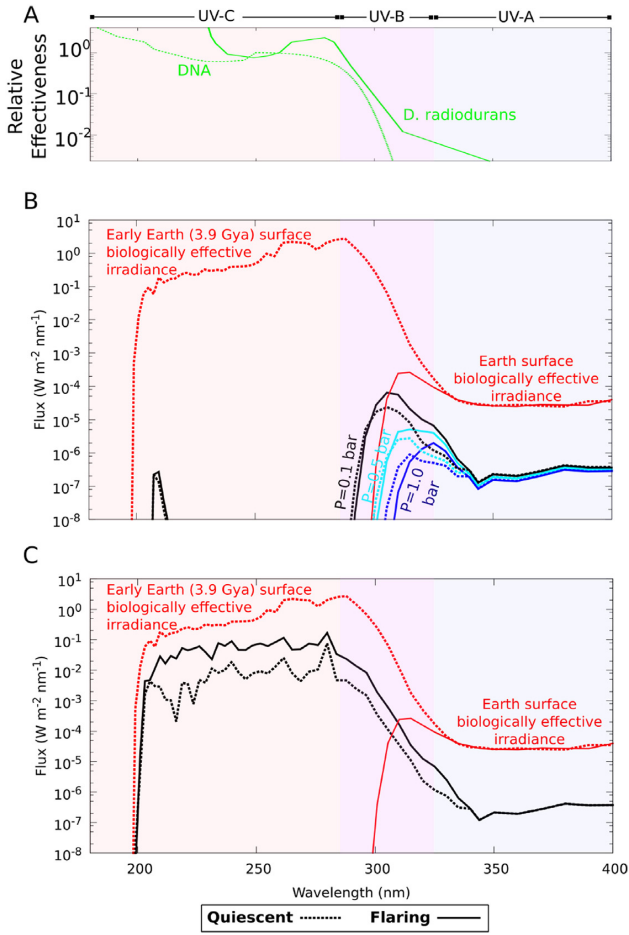

**Figure 4.** Relative biological effectiveness of UV surface radiation on Proxima-b. (A) The biological effectiveness of UV on DNA and the radiation-resistant microorganism D. radiodurans (Voet et al. 1963; Diffey 1991) quantifies the relative effectiveness of different wavelengths of UV radiation to cause DNA destruction or, for D. radiodurans, mortality, which increases with decreasing wavelength. Biological effectiveness of UV damage for (B) oxygenic atmospheres and (C) anoxic atmosphere models shown as convolution of the surface UV flux and action spectrum over wavelength (solid line shows flaring, dashed line quiescent star), compared to present-day Earth (red solid) and early Earth (3.9 billion years ago) (red dashed).

Fig. 4 shows that only for highly eroded (0.1 bar) and anoxic atmosphere models the planet's biologically effective UV radiation surface flux is higher than for present-day Earth. While the anoxic atmosphere does result in a considerably more biologically harmful radiation environment compared to the present-day Earth, it is still approximately an order of magnitude less biologically harmful than early Earth's. Therefore, UV surface radiation levels should not rule out surface habitability for our closest potentially habitable planets or for planets orbiting in the HZ of active M stars in general. Furthermore, in the absence of an ozone layer, depending on the atmospheric composition of a planet, other atmospheric gases, such as sulfur compounds or $CO_2$ can absorb UV radiation (see e.g. Cockell et al. 2000a; Rugheimer et al. 2015). The production of organic hazes via methane photochemistry, which may have existed in the early Earth's anoxic atmosphere, could also act as a UV absorber for planets that lack significant ozone (Arney et al. 2017). This further strengthens the argument to not rule out our neighbouring HZ planets in the search for life.

### 3.3 UV can create as well as hide signatures for life

For late-type M stars, high UV surface environments could remain for billions of years (see e.g. West et al. 2004; France et al. 2013; Rugheimer et al. 2015; Youngblood et al. 2016). In a high UV surface environment, mechanisms that protect biota from such radiation can play a crucial role in maintaining surface habitability, especially on planets around active M stars with thin, eroded or anoxic atmospheres, where other UV-attenuating gases/particles are not present.

On Earth, biological mechanisms such as protective pigments and DNA repair pathways (see e.g. Neale & Thomas 2016) or biofluorescence (O'Malley-James & Kaltenegger 2018) can prevent, or mitigate, radiation damage. Some microorganisms and lichens have been observed to tolerate full solar UV in space exposure experiments, often using protective cells or pigments as UV-screens (e.g. Cockell 1998; Sancho et al. 2007; Onofri et al. 2012). Strategies such as living under a soil/sand layer, in rock crevices, and under water could also be used (e.g. Cockell et al. 2000a; Cockell, Kaltenegger & Raven 2009; Ranjan & Sasselov 2016). Only 1 μm of water is needed to attenuate the shortest UV-C wavelengths (<168 nm) by a factor of 10 or more (O'Malley-James & Kaltenegger 2017). If small insoluble particles are present in a water column, scattering could effectively reduce the UV-C flux by ∼40 per cent at a depth of just 1 cm (Cockell, Southern & Herrera 2000b). Survival strategies like these could lead to a cryptic surface biosphere that produces no detectable surface biosignatures (Cockell et al. 2009). However life in subsurface environments should produce no, or very weak, surface biosignatures.

Alternative UV protective mechanisms like biofluorescence (see O'Malley-James & Kaltenegger 2018 for details) could indicate life remotely. Fluorescence is common in the natural world and in some cases may serve as protection against UV radiation damage by upshifting UV light to longer, safer wavelengths (O'Malley-James & Kaltenegger 2018).

Protective biofluorescence would be most useful during the increase in UV flux during flares, and could cause a temporary change in the planet's surface brightness in the visible. Constant high UV radiation environment present in the anoxic planet model could favour continuous fluorescence (O'Malley-James & Kaltenegger 2018). Lab experiments with green fluorescent proteins have successfully produced high fluorescent efficiencies of up to 100 per cent (see O'Malley-James & Kaltenegger 2018 and references therein). Because biofluorescence is independent of the visible flux of the host star and only dependent on the UV flux of the star, emitted biofluorescence can increase the visible flux of a planet orbiting an active M-star by several orders of magnitude (O'Malley-James & Kaltenegger 2018) during a flare.

Detection of the 9.6 μm ozone band could provide the first insights into UV surface environments of planets in nearby M star systems, showing whether or not those atmospheres contain ozone. Ozone and other atmospheric gases are potentially detectable by near future telescopes like the James Webb Space Telescope for our closest planet, Proxima-b (Kreidberg & Loeb 2016), while high-contrast imaging with planned ground-based telescopes like the ELT could provide additional atmospheric characterization (see e.g. Lovis et al. 2017; Luger et al. 2017; Snellen et al. 2017).

## 4 CONCLUSIONS

While a multitude of factors ultimately determine an individual planet's habitability our results demonstrate that high UV radiation levels may not be a limiting factor. The compositions of the atmospheres of our nearest habitable exoplanets are currently unknown; however, if the atmospheres of these worlds resemble the composition of Earth's atmosphere through geological time, UV surface radiation would not be a limiting factor to the ability of these planets to host life. Even for planets with eroded or anoxic atmospheres orbiting active, flaring M stars the surface UV radiation in our models remains below that of the early Earth for all cases modelled. Therefore, rather than ruling these worlds out in our search for life, they provide an intriguing environment for the search for life and even for searching for alternative biosignatures that could exist under high-UV surface conditions.

## ACKNOWLEDGEMENTS

The authors acknowledge funding from the Simons Foundation (290357, Kaltenegger). We thank the anonymous referee for helpful comments and suggestions.

## REFERENCES

Airapetian V. S., Glocer A., Khazanov G. V., Loyd R. O. P., France K., Sojka J., Danchi W. C., Liemohn M. W., 2017, ApJ, 836, L3
Anglada-Escudé G. et al., 2016, Nature, 536, 437
Arney G. N., Meadows V. S., Domagal-Goldman S. D., Deming D., Robinson T. D., Tovar G., Wolf E. T., Schwieterman E., 2017, ApJ, 836, 49
Barnes R. et al., 2018, preprint (arXiv:1608.06919)
Calkins J., Barcelo J. A., 1982, The Role of Solar Ultraviolet Radiation in Marine Ecosystems. Springer, US, p. 143
Cockell C. S., 1998, Theor. Biol., 193, 717
Cockell C. S., Catling D. C., Davis W. L., Snook K., Kepner R. L., Lee P., McKay C. P., 2000a, Icarus, 146, 343
Cockell C. S., Southern A., Herrera A., 2000b, Ecol. Eng., 16, 293
Cockell C. S., Kaltenegger L., Raven J. A., 2009, Astrobiol., 9, 623
Davenport J. R. A., Kipping D. M., Sasselov D., Matthews J. M., Cameron C., 2016, ApJ, 829, L31
Diffey B. L., 1991, Phys. Med. Biol., 36, 299
Dittmann J. A. et al., 2017, Nature, 544, 333
Dong C., Lingam M., Ma Y., Cohen O., 2017, ApJ, 837, L26
Dressing C. D., Charbonneau D., 2013, ApJ, 767, 95
Dressing C. D., Charbonneau D., 2015, ApJ, 807, 45
France K. et al., 2013, ApJ, 763, 149
France K. et al., 2016, ApJ, 820, 89
Gaidos G., 2013, ApJ, 770, 90
Garcia-Sage K., Glocer A., Drake J. J., Gronoff G., Cohen O., 2017, ApJ, 844, L13
Garraffo C., Drake J. J., Cohen O., 2016, ApJ, 833, L4
Gillon M. et al., 2016, Nature, 533, 221
Goldblatt C., 2018, preprint (arxiv:1608.07263)
Kaltenegger L., 2017, ARA&A, 55, 433
Kaltenegger L., Traub W. A., Jucks K. W., 2007, ApJ, 658, 598
Kasting J. F., Holland H. D., Pinto J. P., 1985, J. Geophys. Res., 90, 10497
Kerwin B. A., Remmele R. L., 2007, J. Pharm. Sci., 96, 1468
Kopparapu R., Wolf E. T., Arney G., Batalha N. E., Haqq-Misra J., Grimm S. L., Heng K., 2017, ApJ, 845, 5
Kreidberg L., Loeb A., 2016, ApJ, 832, L12
Lingam M., Loeb A., 2017, ApJ, 846, L21
Lovis C. et al., 2017, A&A, 599, A16
Loyd R. P., Shkolnik E. L., Schneider A. C., Barman T. S., Meadows V. S., Pagano I., Peacock S., 2018, ApJ, 867, 70
Luger R., Lustig-Yaeger J., Fleming D. P., Tilley M. A., Agol E., Meadows V. S., Deitrick R., Barnes R., 2017, ApJ, 837, 63
Meadows V. S. et al., 2018, Astrobiol., 18, 133
Neale P. J., Thomas B. C., 2016, Astrobiol., 16, 245
O'Malley-James J. T., Kaltenegger L., 2017, MNRAS, 469, L26
O'Malley-James J. T., Kaltenegger L., 2018, MNRAS, 481, 2487
Onofri S. et al., 2012, Astrobiol., 12, 508
Pavlenko Y., Mascareño A. S., Rebolo R., Lodieu N., Béjar V. J. S., Hernández J. G., 2017, 606, A49
Ramirez R. M., Kaltenegger L., 2014, ApJ, 797, L25
Ranjan S., Sasselov D. D., 2016, Astrobiol., 16, 68
Ribas I. et al., 2016, A&A, 596, A111
Rimmer P. B., Xu J., Thompson S. J., Gillen E., Sutherland J. D., Queloz D., 2018, Sci. Adv., 4, eaar3302
Rothschild L. J., Mancinelli R. L., 2001, Nature, 409, 1092
Rugheimer S., Segura A., Kaltenegger L., Sasselov D., 2015, ApJ, 806, 137
Sancho L. G., de la Torre R., Horneck G., Ascaso C., de los Rios A., Pintado A., Wierzchos J., Schuster M., 2007, Astrobiol., 7, 443
Segura A., Kasting J. F., Meadows V., Cohen M., Scalo J., Crisp D., Butler R. A. H., Tinetti G., 2005, Astrobiol., 5, 706
Segura A., Walkowicz L. M., Meadows V., Kasting J., Hawley S., 2010, Astrobiol., 10, 751
Setlow J. K., Boling M. E., 1965, Biochimica et Biophysica Acta (BBA)-Nucleic Acids and Protein Synthesis, 108, 259
Snellen I. A. G. et al., 2017, AJ, 154, 77
Turbet M., Leconte J., Selsis F., Bolmont E., Forget F., Ribas I., Raymond S. N., Anglada-Escudé G., 2016, A&A, 596, A112
Vidotto A. A., Jardine M., Morin J., Donati J.-F., Lang P., Russell A. J. B., 2013, A&A, 557, A67
Voet D., Gratzer W. B., Cox R. A., Doty P., 1963, Biopolym., 1, 193
West A. A. et al., 2004, ApJ, 128, 426
Youngblood A. et al., 2016, ApJ, 824, 101
Bonfils X.et al., 2018, Astron. Astrophys., 613, A25
Scalo J.et al., 2007, Astrobiology, 7, 85
Tarter J. C.et al., 2007, Astrobiology, 7, 30
Shields A.L. , Ballard S., Johnson J. A., 2016, Physics Reports, 663, 1
Lammer H.et al., 2007, Astrobiology, 7, 185
See V.et al.,2014, Astron. Astrophys., 570, A99
Buccino A. P. , Lemarchand G. A., Mauas P. J., 2006, Icarus, 183, 491
Buccino A. P., Lemarchand G. A., Mauas P. J., 2007, Icarus, 192, 582
West A. A.et al., 2011, Astronom. J., 141, 97
Tilley M. A., Segura A., Meadows V. S., Hawley S., Davenport J., 2019, Astrobiology, 19, 64
Heath M. J., Doyle L. R., Joshi M. M., Haberle R. M., 1999, Origins of Life and Evolution of the Biosphere, 29, 405
Bazot M., Christensen-Dalsgaard J., Gizon L., Benomar O., 2016, MNRAS, 460, 1254
Kaltenegger L., Sasselov D., 2009, Astrophys. J. , 708, 1162
Toon O. B., McKay C. P., Ackerman T. P., Santhanam K., 1989, J. Geophys. Res.: Atmospheres, 94, 16287
Pavlov A. A., Kasting J. F., Brown L. L., Rages K. A., Freedman R., 2000, J. Geophys. Res., 105, 981
Kopparapu R. K., Ramirez R. M., SchottelKotte J., Kasting J. F., Domagal-Goldman S., Eymet V., 2014, Astrophys. J. Lett., 787, L29
Segura A., Meadows V. S., Kasting J. F., Crisp D., Cohen M., 2007, Astron. Astrophys., 472, 665
Kharecha P., Kasting J. F., Siefert J., 2005, Geobiology, 3, 53